\documentclass[showpacs,pre,amsmath,amssymb,superscriptaddress]{revtex4}
\usepackage{graphicx}
\usepackage{amsmath}
\usepackage{amssymb}
\usepackage{units}
\newcommand{\kon}{k_\textrm{\scriptsize{on}}c_0}
\newcommand{\koff}{k_\textrm{\scriptsize{off}}}

\newcommand{\SA}{\alpha}
\newcommand{\SI}{\beta}
\newcommand{\DA}{\gamma}
\newcommand{\DI}{\delta}

\begin{document}
\title{Noise Effects  in Nonlinear Biochemical Signaling}
 \author{Neda Bostani}
\affiliation{Key Laboratory of Particle Astrophysics, Institute of High Energy Physics, Chinese Academy of Sciences, Beijing 100049, China}
\author{David A. Kessler}
 \email{kessler@dave.ph.biu.ac.il}
\affiliation{Department of Physics, Bar-Ilan University,
 Ramat-Gan 52900 Israel }
 \author{Nadav M. Shnerb}
 \email{shnerbn@mail.ph.biu.ac.il}
\affiliation{Department of Physics, Bar-Ilan University,
 Ramat-Gan 52900 Israel }
\author{Wouter-Jan Rappel}
\email{rappel@physics.ucsd.edu}
\affiliation{Center for Theoretical Biological Physics,
University of California San Diego, La Jolla, CA 92093-0319 USA }
\author{Herbert Levine}
\email{hlevine@ucsd.edu}
\affiliation{Center for Theoretical Biological Physics,
University of California San Diego, La Jolla, CA 92093-0319 USA }

\begin{abstract}
It has been generally recognized that stochasticity can play an important role in the information processing accomplished by reaction networks in biological cells. Most treatments of that stochasticity employ Gaussian noise even though it is a priori obvious that this approximation can violate physical constraints, such as the positivity of chemical concentrations. Here, we show that even when such nonphysical fluctuations are rare, an exact solution of the Gaussian model shows that the model can yield unphysical results. This is done in the context of a simple incoherent-feedforward model which exhibits perfect adaptation in the deterministic limit. We show how one can use the natural separation of time scales in this model to yield an  approximate model, that is 
analytically solvable,
including its dynamical response to an environmental change.  Alternatively, one can employ a cutoff procedure to regularize the Gaussian result.

\end{abstract}

\pacs{02.50.Le, 05.65.+b, 87.23.Ge, 87.23.Kg}

\maketitle
\section{Introduction}
The role of stochasticity in the functioning of cellular signal transduction networks is a question of great topical interest~\cite{stochastic}. Unlike typical condensed-matter systems, biological cells must carry out chemical manipulations with small numbers of molecules, an inherently noisy situation. Noise comes in a variety of forms, including fluctuations in chemicals to be sensed~\cite{berg-purcell}, fluctuations in the binding-unbinding of receptor arrays~\cite{hu-prl}, fluctuations during the processing of information~\cite{bialek}, and fluctuations in the implementation of downstream actions~\cite{yuhai-prl}.

In this context, almost all analytic studies of stochastic reaction dynamics utilize a small noise Gaussian approximation. This assumption emerges naturally, for example, in the van Kampen system-size expression~\cite{van-kampen} where the fluctuations in particle number are formally lower order and hence are treated as small and centered around the mean value set by the deterministic reaction equations. The initial purpose of this paper is to point out that this approach may give highly misleading results especially when some of the downstream reactions are nonlinear. We do this by studying a specific example, that of an incoherent feedforward module processing data from a small number of receptors~\cite{tang}. Afterwards, we show how an alternate approach for the rapid activation versus slow inhibition limit can provide a complementary analytic approach.

\begin{figure}
\includegraphics[width=0.7\textwidth]{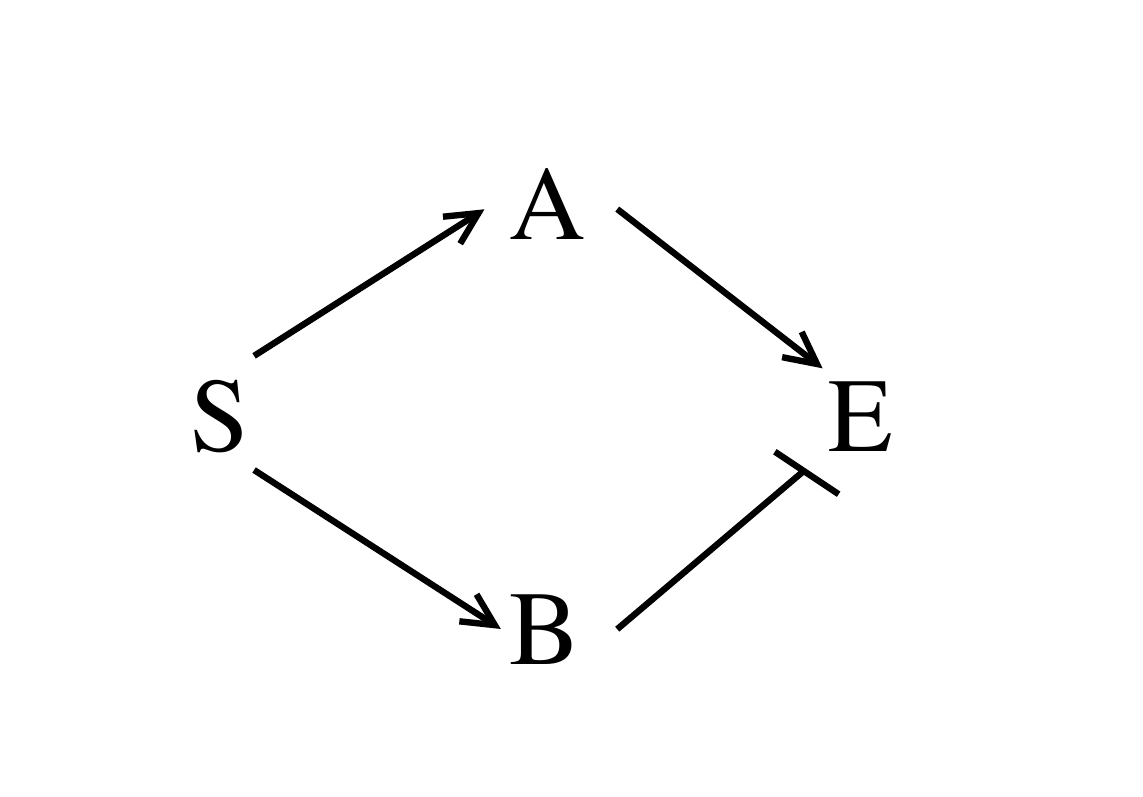}
\caption{Circuit diagram showing the activation of both $A$ and $B$ by the signal $S$.  $A$ in turn activates $E$, while $B$ inhibits it.}
\label{circuit}
\end{figure}

The example we choose to study is a part of the gradient sensing module underlying the chemotactic response of Dictyostelium cells~\cite{devreotes}.  These cells appear to implement a control circuit incorporating a simple incoherent feedforward loop topology for adapting out the constant concentration background~\cite{iglesias,koske}. This circuit is instantiated by using a RAS-GEF as a positive signal and RAS-GAP as the complementary inhibitor~\cite{ras-footnote}. These are both activated upon the binding of cAMP by the G-protein coupled cAMP receptor and in turn  drive the signaling hub protein RAS into its active (GTP-bound), respectively inactive (GDP-bound) form. The circuit diagram is illustrated in Fig. \ref{circuit}.  In the limit where we are far from saturation, the system can be described by the following equations~\cite{koske}:
\begin{eqnarray}
\frac{dA}{dt} & = & \SA S(t) - \DA A \nonumber \\
\frac{dB}{dt} & = & \SI S(t) - \DI B \nonumber \\
\frac{dE}{dt} & = & A (1-E) - BE 
\end{eqnarray}
Here $E$ is the fraction of RAS molecules that have been activated and $S$ is the external signal which drives both the activator $A$ and inhibitor $B$. It is trivial to verify that if $S$ is a constant signal, the steady-state value of $E$, $E_0=\SA\DI/(\SA\DI + \SI\DA)$ does not depend on its value. Hence this system in the deterministic limit is a perfectly adapting module, exhibiting only a transient response to changes in its input.

In order to study the effect of noise in the input signal $S(t)$ on this system, it is standard to assume that $S(t)$ is the sum of a deterministic signal $S_0(t)$ plus stationary Gaussian noise.  For simplicity, we consider the case where the deterministic signal is a constant, $S_0$, and the noise, $\eta(t)$, has zero mean and
correlator
\begin{equation}
\langle \eta(t) \eta(t') \rangle = \sigma^2 e^{-|t-t'|/\tau}
\end{equation}
The advantage of assuming Gaussian noise is that the system is then analytically tractable.  The fields $A$ and $B$ can be expressed in term of the noise
$\eta$ as follows:
\begin{eqnarray}
A(t) & = \frac{\SA}{\DA} S_0 + \SA & \int _{-\infty} ^t dt' \ e^{-\DA (t-t')} \eta(t') \nonumber \\
B(t) & = \frac{\SI}{\DI}S_0 +  \SI & \int _{-\infty} ^t dt' \ e^{-\DI (t-t')} \eta(t') 
\end{eqnarray}
Substituting this into the effector equation allows us to find
\begin{equation}
E(t) = \int_{-\infty} ^t dt_1 \ g_1 (t_1) e^{-\int_{t_1}^{t} dt_2 \ g_2(t_2)}
\end{equation}
with the definitions
\begin{eqnarray} 
g_1( t) & = &  \frac{\SA}{\DA}S_0 + \int_{-\infty}^t dt' \ \SA e^{-\DA (t-t')} \eta(t')  \nonumber \\
g_2( t) & = &  \left(\frac{\SA}{\DA} + \frac{\SI}{\DI}\right) S_0 + \int_{-\infty}^t dt' \ \left( \SA e^{-\DA (t-t')} +\SI e^{-\DI (t-t')} \right)  \eta(t') 
\end{eqnarray}
From this expression, all moments of $E$ can be calculated exactly. For example, let us focus on the expectation values of E. The standard expressions for Gaussian processes, for example
\begin{equation}
\langle e^ {-\int (S(t')- s_\textit{\tiny{eq}}) h(t') dt'}\rangle = e ^{\frac{1}{2} \int dt' dt'' h(t') G(t',t'') h(t'')} 
\end{equation} 
allow us after a tedious calculation to derive the following expression for $\langle E\rangle$:
\begin{equation}
\langle E \rangle  = \int _0^\infty du \ e^{-S_0 \left( \frac{\SA}{\DA} +\frac{\SI}{\DI} \right) u} e^{\sigma^2 \Phi (u) } \left( \frac{\SA S_0}{\DA}-\SA \sigma^2 \Delta _E (u) \right) \label{mean-eq}
\end{equation}
From this, one can immediately recover the aforementioned $\sigma^2 =0$ deterministic result, $\langle E \rangle = \frac{\DI \SA}{\DI \SA +\SI \DA}$. The exact expression for $\Delta_E(u)$, given in the appendix, is not particularly informative; the only critical feature is that it decays to a constant at large $u$. The factor $\sigma^2 \Phi (u)$ is the exponent is much more significant. Again, we leave the full form for the Appendix, and merely give the large $u$ behavior:
\begin{equation}
\sigma^2 \Phi (u)  \sim u \sigma^2 \tau \left( \frac{\SA}{\DA} +\frac{\SI}{\DI} \right) ^2
\end{equation}
This term represent the diffusive growth of (the integral of) $S^2$ and has a well-defined form even in the white-noise limit for $S$ where $\tau \rightarrow 0$ with $\sigma^2 \tau$ fixed. 

\begin{figure}
\hspace*{-.3in}\includegraphics[width=0.48\textwidth]{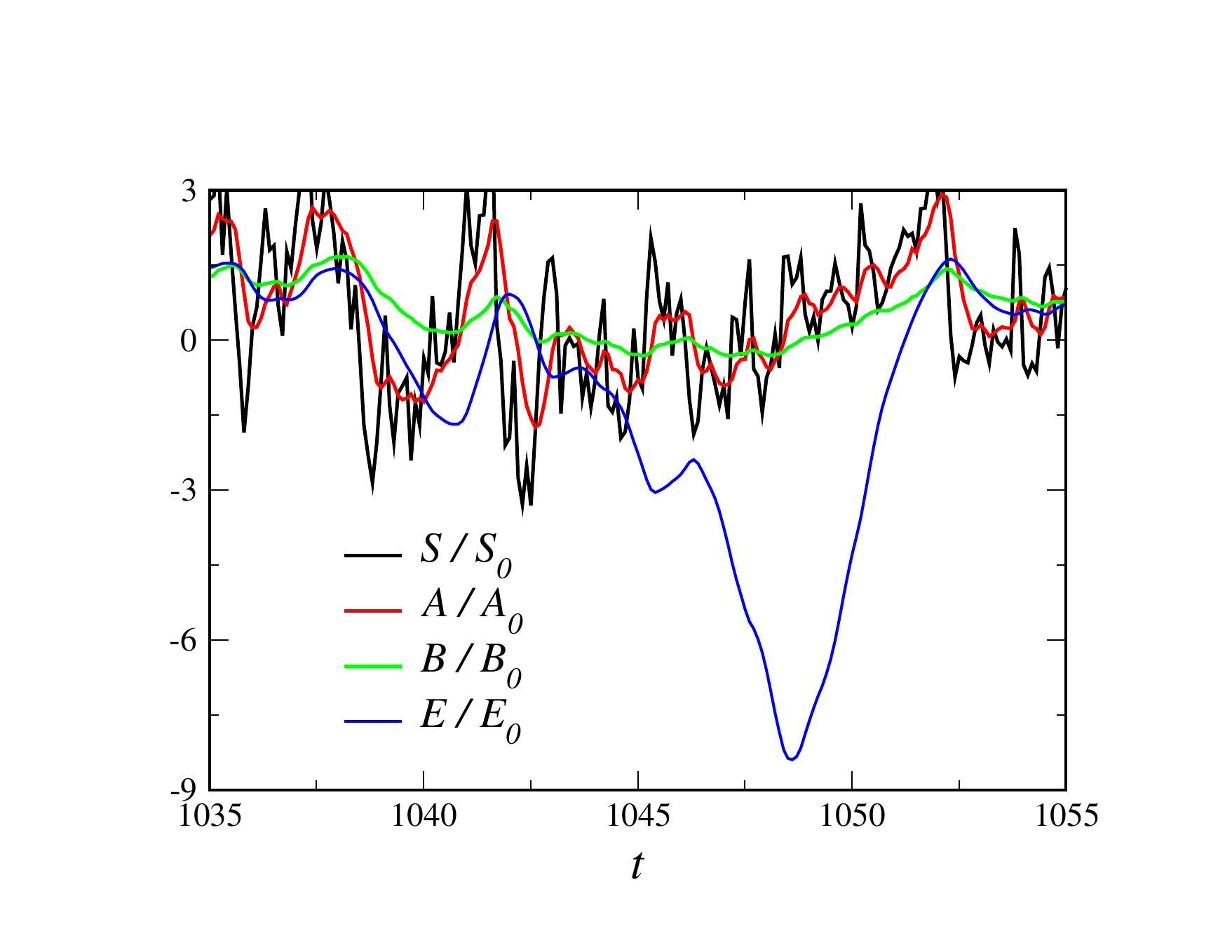}\hspace*{-0.3in}
\includegraphics[width=0.48\textwidth]{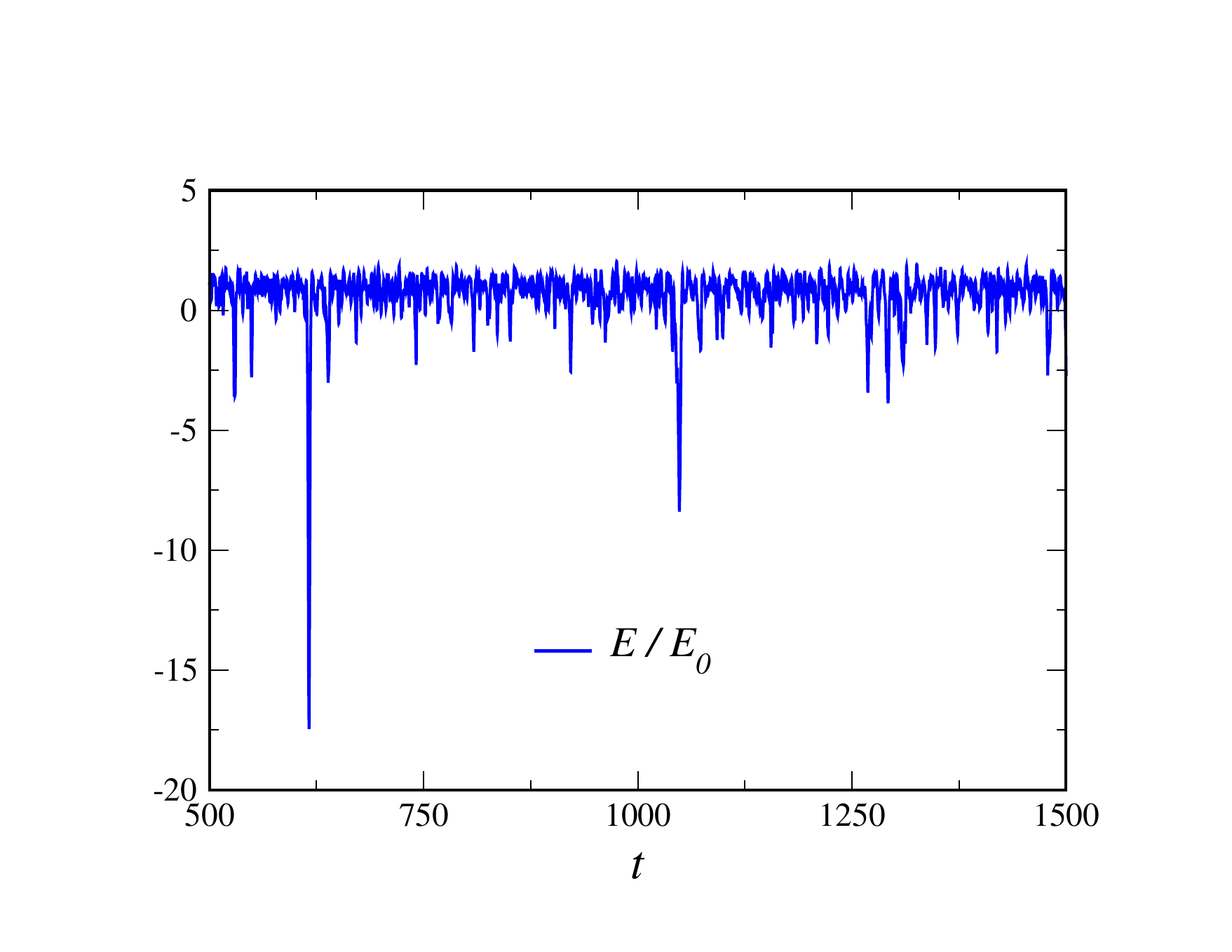}
\caption{(color online). Left) Excerpt of a simulation of the Gaussian  model with $S_0=1/3$, $\sigma^2=2/9$, $\SA=1$, $\DA =2.5$, $\DI = 0.4$, $\SI =1.7$,
showing a large negative fluctuation of $E$. Note that $\SI$ is chosen to be exactly the critical value at which Eq. (\ref{critical}) is violated and $\langle E \rangle$ diverges,  Also shown are
$S(t)/S_0$, $A(t)/A_0$ and $B(t)/B_0$. Right) Larger time series of $E(t)$ showing the intermittent nature of the large fluctuations.}
\label{traces}
\end{figure}

The starting point of our work is the observation that the integral defining $\langle E\rangle$ fails to converge unless
\begin{equation}
S_0 > \sigma^2 \tau \left( \frac{\SA}{\DA} +\frac{\SI}{\DI} \right)
\label{critical}\end{equation}
It is easy to show that similar, but more stringent, bounds hold for all moments of $E$, which therefore are predicted to grow without bound (starting from any initial state) if the noise is too large. The problem arises from the fact that the $S$ fluctuations are unbounded below for Gaussian noise and hence can drive $A$ and/or $B$ negative. This gives rise to transient periods during which $|E|$ {\em grows} exponentially. This behavior can be directly seen in a simulation of the Gaussian noise model, as presented in the left panel of Fig. \ref{traces}. (In passing, these large excursions are rare events if $\sigma$ is sufficiently small,  and hence getting an accurate measure of $\langle E \rangle$ from the simulations can be difficult. We will return to this point in more detail later). As the noise gets large these growth periods lead to ever increasing values of $|E|$ and the anomalous contributions to $\langle E \rangle $ never saturate. 
It is important to note that  even when $S_0/\sigma \ll 1$, so that the negative fluctuations of $A$ and $B$ are rare, nevertheless the Gaussian model can yield unphysical, infinite results, if $\SA/\DA + \SI/\DI$ is large enough.  Thus, our finding depends essentially on the nonlinear coupling of the noisy signal to the $E$ field, so that the noise acts  multiplicatively on $E$.

\section{Binomial Noise}
Clearly, the Gaussian noise approach is in general unacceptable.  We must treat the noise in a more realistic fashion if we are to have a well-defined model.  If the source
of the noise is the finite number of receptors~\cite{rappel-pnas,wingreen}, we are led to consider a model wherein the incoming signal is a random variable reflecting the fraction of bound sensors. 
Assuming that the $N$ receptors are independent and bind a ligand of fixed concentration, $c_0$, the signal can be exactly described via the master equation for the probability distribution, $P(s,t)$, for the number $s=0,1\ldots,N$ of occupied receptors. where the signal $S=s/N$; i.e., the fraction of occupied receptors.
  \begin{eqnarray}
\frac{\partial P(s,t)}{\partial t} &=&    - \big[ \koff s   +\kon  (N-s) \big] P(s,t) +  \koff(s+1)P(s+1,t) \nonumber \\  &\ &{}+  
 \kon  (N-s+1) P(s-1,t)  \nonumber \label{master}\end{eqnarray}
 We call the model with this discrete noise the binomial noise model, as the
 equilibrium distribution of $s$ is binomial,
\begin{equation}
P_{eq} (s) = \frac{N!}{s!(N-s)!} \frac{ (\kon ) ^s (\koff)^{N-s}}{(\kon + \koff )^N} \ .
\end{equation}

\begin{figure}
\includegraphics[width=0.7\textwidth]{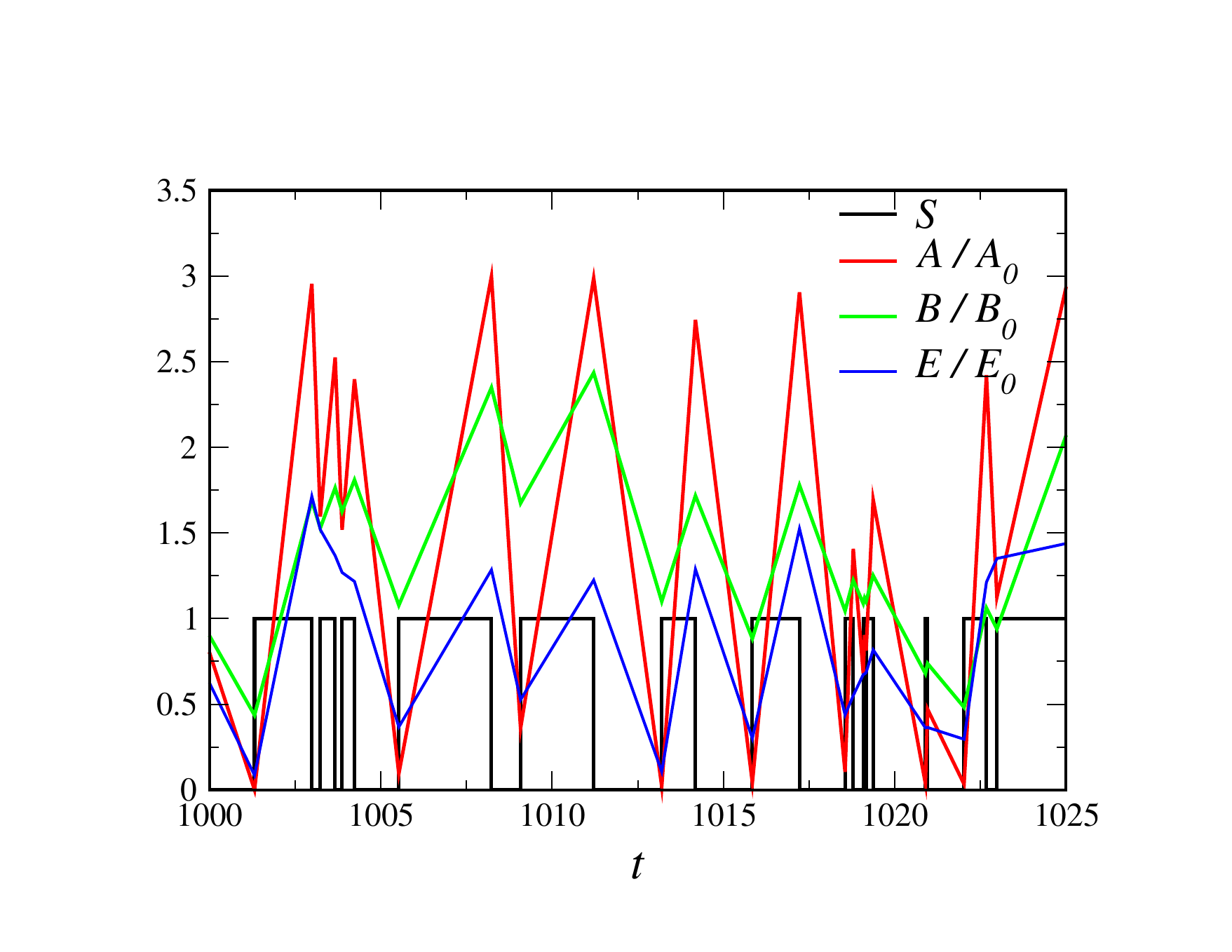}
\caption{(color online).  Excerpt of a simulation of the $N=1$ binomial  model with parameters parallel to those of the Gaussian simulation in Fig. \ref{traces}: $\kon=1$, $\koff=2$, $\SA=1$, $\DA =2.5$, $\DI = 0.4$, $\SI =1.7$. }
\label{trace_bi}
\end{figure}

For large $N$,  we recover a Gaussian noise process, except in the tails, as can be seen via the following argument. Our discrete stochastic  process, Eq. \ref{master}, is  well approximated for large $N$ by the Ornstein-Uhlenbeck process, described by the Fokker-Planck equation
\begin{equation}
\tau \frac{\partial P_G(S,t)}{\partial t} = \frac{\partial}{\partial S} \left[ (S-S_0) P_G  \right] +  \sigma_S^2 \frac{\partial ^2 P_G}{\partial S^2}
\end{equation}
with $S_0=\kon/(\kon + \koff)$, $\tau ^{-1} = \koff +\kon $ and  $\sigma_S^2 =  \koff \kon/( N (\kon+\koff)^2) $ so that in the limit the stochastic function $S(t)$ is Gaussian with variance $\sigma _S$. Here, of course, the equilibrium distribution is
\begin{equation}
P_{G,eq}(S) \simeq \exp{ \left[ - \left(S-S_0\right)^2 / 2 \sigma_S^2 \right] }
\end{equation} and the steady-state autocorrelation function is
\begin{equation} 
< S(t)S(t')> - \ S_0^2 \equiv G(t,t') = \sigma_S^2 e^{-|t-t'|/\tau}
\end{equation}
Nevertheless, for all finite $N$, the signal $S$ in the binomial process is always non-negative, and no anomalous behavior can occur, with $E(t)$ strictly bounded from below by $0$.  The striking difference in the two models is apparent by comparing the Gaussian simulation presented in Fig. \ref{traces} to  the
 simulation of the corresponding binomial model in Fig. \ref{trace_bi}.  The well-behaved nature of the binomial model for all $N$  is consistent with the fact 
that our condition for the convergence of the first moment in the Gaussian model, Eq. (\ref{critical}), is always satisfied in the large $N$ limit, given that the noise amplitude $\sigma_S^2$ is small, of order $O(1/N)$.  The large-$N$ limit, however, is only an asymptotic approximation, since for any finite value of the noise amplitude $\sigma_S^2$ of the Gaussian model, sufficiently high moments of $E$ do indeed diverge.

\begin{figure}
\includegraphics[width=0.7\textwidth]{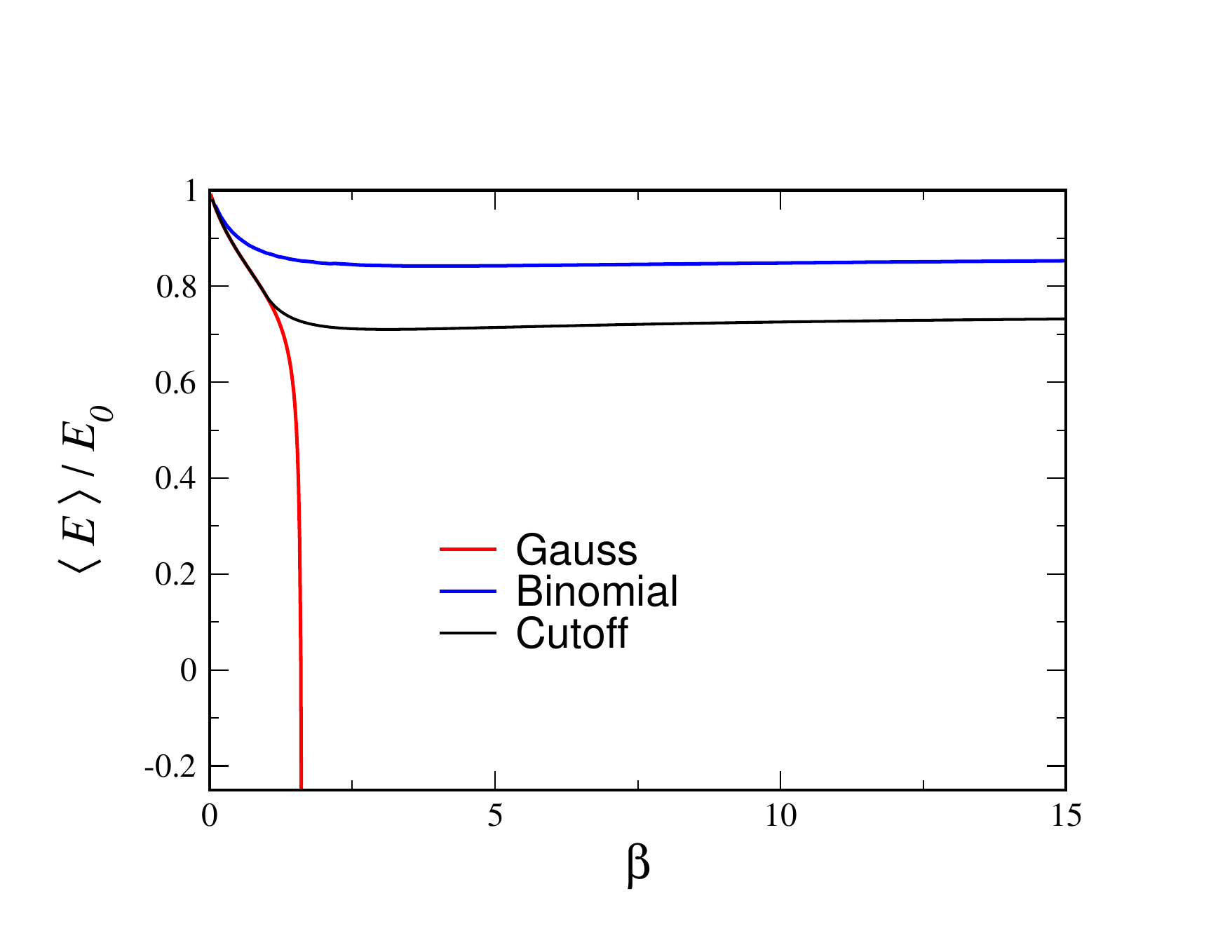}
 \caption{(color online). Variation of $\langle E \rangle$ with $\SI$ for the Gaussian model, Eq. \ref{mean-eq} and the corresponding $N=1$ binomial model, derived from averaging $10^4$ simulations. The parameters of the Gaussian model (except $\SI$) are as  in Fig. \ref{traces}; for the binomial model as in Fig. \ref{trace_bi}.  These results are compared to that of the cutoff model (defined later in the text).}
 \label{comp}
\end{figure}

\begin{figure}
\includegraphics[width=0.7\textwidth]{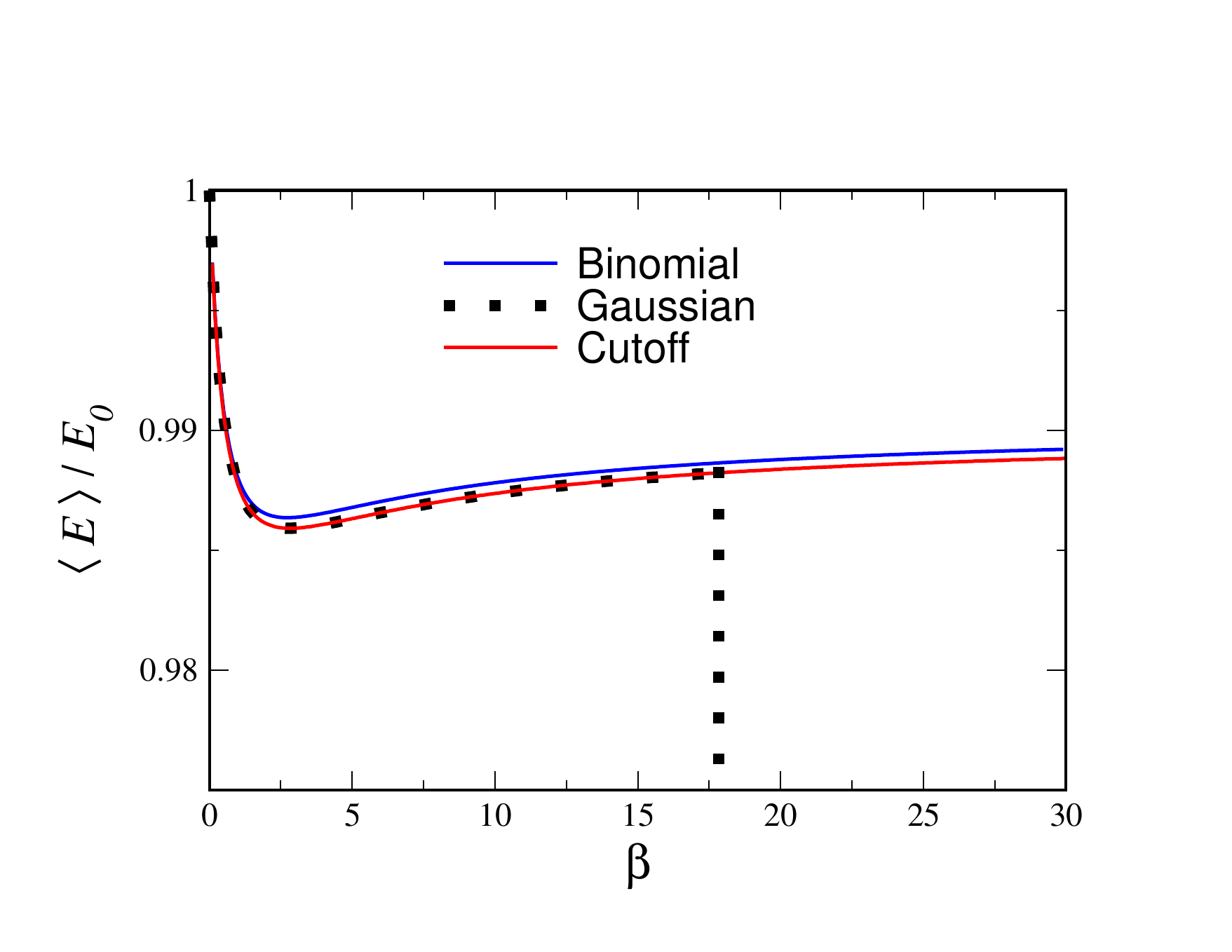}
 \caption{(color online).Variation of $\langle E \rangle$ with $\SI$ for the Gaussian model, Eq. \ref{mean-eq} and the corresponding binomial model, derived from averaging $10^4$ simulations. Here $N=10$ in the binomial model, and similarly the Gaussian model has $\sigma^2=4/90$, so that the Gaussian $\langle E \rangle$ diverges at $\SI = 17.84$.  The other parameters are as  in Fig. \ref{comp}.  These results are compared to that of the cutoff model (defined later in the text).}
\label{comp10}
\end{figure}

An additional perspective on the difference between the binomial and its parallel Gaussian model is afforded by examining the equilibrium $\langle E \rangle$ as a function of the parameter $\SI$.  This is presented in Fig. \ref{comp}.  In the Gaussian noise model, the divergence condition condition is obviously satisfied for $\SI>\SI_c$, since the right-hand size grows linearly with $\SI$. The incipient divergence of $\langle E \rangle$ for the Gaussian model at $\beta_c=1.7$ is apparent.  The binomial model, on the other hand, shows no special behavior at the Gaussian critical $\SI$ at which $\langle E \rangle$ diverges.  Rather, $\langle E \rangle$ exhibits a broad minimum at $\SI\approx 3.9$ and then rises toward unity as $\SI$ increases.  Even before the divergence, the Gaussian model deviates significantly from its binomial counterpoint, since binomial noise is far from Gaussian when $N$ is small. The third curve in this figure results from a cutoff version of the Gaussian model, to be discussed later. For smaller noise, (equivalently, larger $N$ in the binomial model), however, as depicted in Fig. \ref{comp10}, the difference between the two models is small, essentially right up to the divergence, as binomial noise is very well approximated by Gaussian
noise, except in the tails, which are dynamically irrelevant as long as one is a finite distance below the transition.  The onset of the deviation for small noise
in the Gaussian model is extremely close to the critical $\SI$.  For example, $\langle E \rangle$ crosses zero at a distance of the order of $~10^{-18}$ from the critical $\SI$ for the parameters of Fig. \ref{comp10}.

\section{Sudden/Adiabatic (S/A) Approximation}
Unfortunately, the binomial model does not admit an analytic solution for finite $N$ (which is of course the interesting case, since otherwise the effect of the noise is infinitesimal).  We can however make use of the natural ordering of time scales in the problem to construct a solvable limit.  For the circuit to show a significant transient response, it is necessary for the time scale of the $A$ dynamics to be much faster than the $B$ dynamics, otherwise the system adapts
too rapidly to the changing signal and the transient response is aborted.  Furthermore, the time scale
of the $E$ dynamics,  $1/(A_0 + B_0)$, should be intermediate to those of the $A$ and $B$ fields. If the $E$ dynamics is too fast, then the system is limited in any case to respond no quicker than $A$, and if  the $E$ dynamics is too slow, $A$ and $B$ have reequilibrated by the time $E$ starts to respond, and again there is no transient response.   Also, the time scale of the noise dynamics should be intermediate.  Too fast noise would just get averaged away, and too slow noise would be adapted away.  Thus, we are led to consider the limit where the $A$ dynamics is much faster than all the other processes, and the $B$ dynamics much slower.  This limit is analytically treatable, as we now proceed to show.

Formally, we define our approximate theory, which we denote the Sudden/Adiabatic (S/A) theory, by taking both $\SI \rightarrow 0$ and $\DI \rightarrow 0$, with 
a fixed ratio $B_p = (\SI /\DI)$. Since the time scale of the $B$ dynamics is so long, the noise in the signal is completely averaged over and we can just set $B=B_0=B_p(\kon/(\kon+\koff))$, its average value. In addition, we  take the limit of large $\SA$ and $\DA$ (with fixed ratio $A_p$) which guarantees that the activator dynamics is fast enough to precisely follow the noise, i.e. $A(t) = S(t) A_p$. 

\begin{figure}
\includegraphics[width=0.7\textwidth]{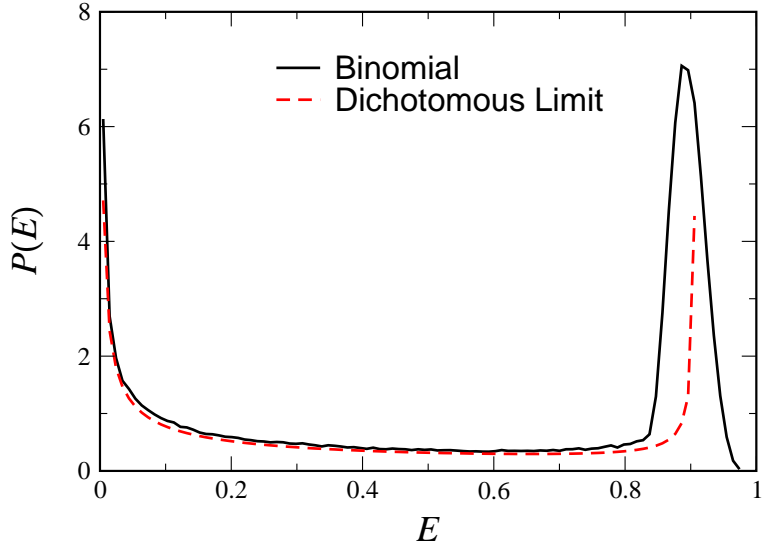}
\caption{The probability distribution function, $P(E)=P_+(E)+P_(E)$ as a function of $E$ for the $N=1$ binomial model with $\SA=100$, $\DA=10$, $\SI=0.2$, $\DI=0.1$, $\kon=\koff=0.4$, compared with the $N=1$ S/A model, with parameters $A_p=10$, $B_0=1$.}
\label{dichot}
\end{figure}

We first examine the case $N=1$.  To proceed, we  decompose the equation for the probability distribution of $E$ into $P_{0,1} (E)$, the joint probability of $E$ and  the input signal $S$ begin 0 or 1, respectively, so that  $P(E,t)=P_0(E,t)+P_1(E,t)$.  We immediately derive that, in steady state,
\begin{eqnarray}
\frac{\partial P_1}{\partial t} & = & \kon P_0 - \koff P_1  + \frac{\partial}{\partial E} \left( \left[  (B_0+A_p)  E -A_p  \right] P_1\right) = 0\nonumber \\
\frac{\partial P_0}{\partial t} & = & - \kon P_0 + \koff P_1 + \frac{\partial}{\partial E} \left(  B_0  E   P_0 \right) = 0
\label{dichoteq}
\end{eqnarray}
Adding the two equations and integrating gives
\begin{equation}
\frac{E_p - E}{B_0}\  P_1 (E)  = \frac{E}{A_p+ B_0} \ P_0 (E)
\end{equation} where $E_p \equiv A_p/(A_p+B_0)$. Substituting this back into Eqs. \ref{dichoteq} and defining $r_+ = \koff / (A_p +B_0)$, $r_- = \kon/B_0$, we can find the normalized probabilities defined on the interval $0 < E < E_p$, 
\begin{eqnarray}
P_1&=& \frac{\Gamma(r_+ +r_- +1)}{\Gamma (r_+) \Gamma (r_-) } E_p^{- (r_++r_-)} \frac{B_0}{\kon+\koff} E^{r_-} \left( E_p - E \right) ^{r_+ -1} \nonumber \\
P_0&=& \frac{\Gamma(r_+ +r_- +1)}{\Gamma (r_+) \Gamma (r_-) } E_p^{- (r_++r_-)} \frac{A_p+B_0}{\kon+\koff} E^{r_- -1} \left( E_p - E \right) ^{r_+ }  
\end{eqnarray}
The total probability $P(E)$  has the interesting behavior of switching from being peaked at the interval center to the interval endpoints, as the parameters are varied~\cite{Ohkubo}.  For $r_+<1$, the probability density diverges at $E_p$, and for $r_-<1$, the density diverges at 0.
The above expression immediately gives the prediction
\begin{equation}
\langle E \rangle =  E_p \frac{r_- + \kon/(\kon+\koff)} {r_++r_-+1}
\end{equation}
To compare this to the full binomial model, we conducted a simulation with $\SA=100$, $\DA=10$ (giving $A_p =10$), $\SI=0.2$, $\DI=0.1$, and $\kon=\koff=0.4$, (giving $B_0=1$). Here, the above theory predicts $\langle E \rangle/E_0= .684$, where the deterministic $E_0=\kon A_p/(\kon A_p + (\kon+\koff)B_0)$.  The simulation gave $\langle E \rangle/E_0=.691$ and indeed the histogram is peaked at the endpoints, as predicted by the analysis.  This is seen in
Fig. \ref{dichot}.

In particular, our limiting theory predicts that as a function of $B_0$, i.e. $\SI$,  $\langle E \rangle / E_0$ starts at a value of unity at $B_0=0$, which is reasonable since the system is saturated and so $\langle E \rangle$ is unity independent of the noise.  For small $B_0$, $\langle E \rangle/E_0$  falls  with an
an initial slope of $\koff/( \kon (\kon+\koff))$.  As a  function of $B_0$, $\langle E \rangle / E_0$ reaches a minimum at $B_0=\sqrt{A_p\kon}$ and then turns back up, approaching unity at large $B_0$.  This qualitative behavior is in accord with what we saw in Fig. 1, even though there the parameters are far from fulfilling the separation of scales assumed in the analysis.  For $\kon=\koff$, the value of $\langle E \rangle / E_0$ at the minimum is 
\begin{equation}
(\langle E \rangle / E_0)_\textit{\scriptsize{min}} = \frac{ (r+2)^2}{2(r^2 + 2r + 2)} ; \qquad\qquad r\equiv  \sqrt{A_p/\kon}
\end{equation}
This decreases from unity for small $r$ to a value of $1/2$ at large $r$. Thus the larger $A_p/\kon$, the larger the noise-induced relative suppression of $\langle E \rangle$, since the effective noise amplitude increases as $\kon$ decreases.  When $\koff \ll \kon$, the value at the minimum approaches unity, so there is no suppression.  On the other hand, when $\koff \gg \kon$, the value at the minimum approaches $\koff/(\koff + A_p)$, which indicates the maximal suppression occurs at $A_p \gg \koff \gg \kon$. 

\subsection{Moment Equations}
\begin{figure}
\includegraphics[width=0.7\textwidth]{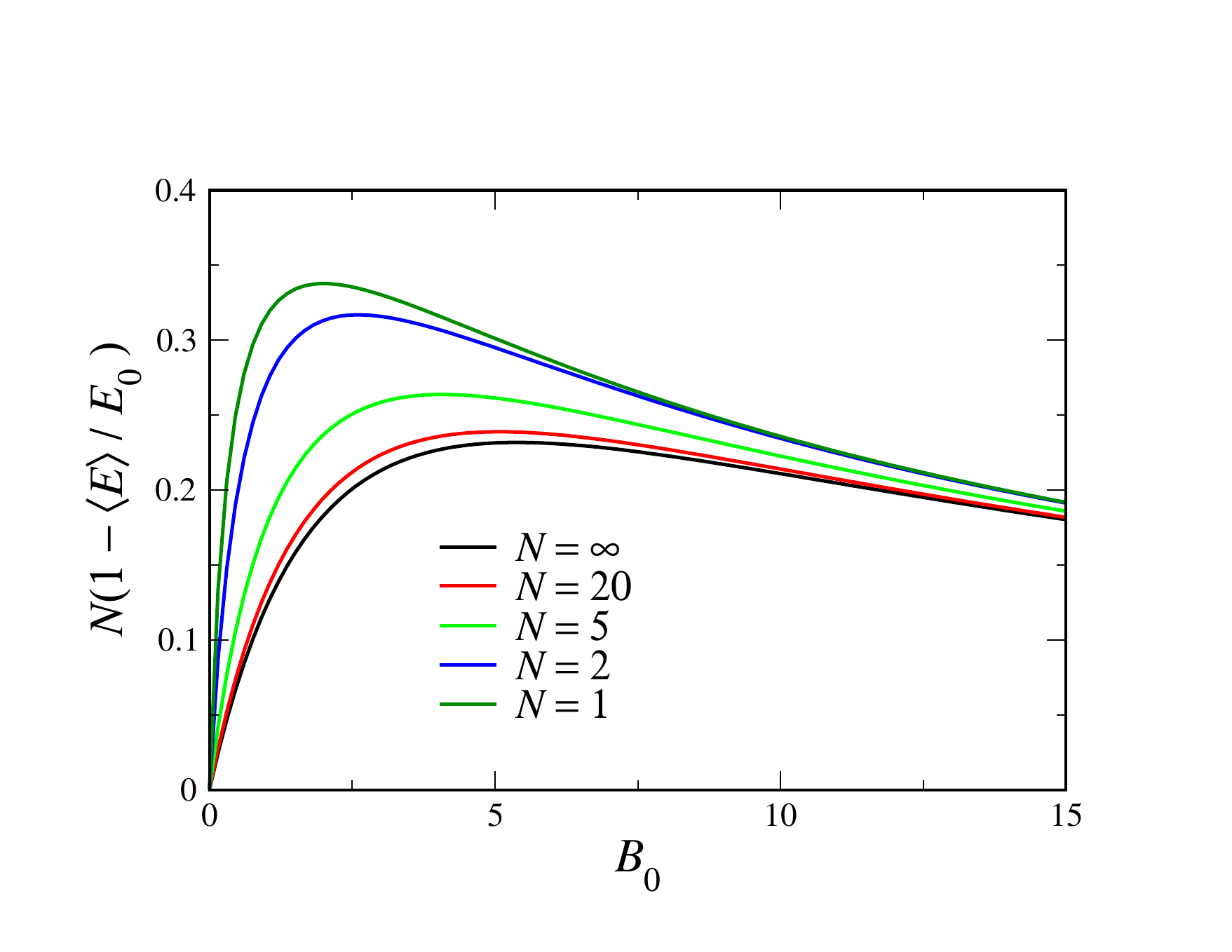}
\caption{The scaled suppression of the equilibrium $\langle E \rangle$, $N(1-\langle E \rangle)$ as a function of $B_0$ in the reduced model, for various $N=1$, $2$, $5$, and $20$ along with the asymptotic large-$N$ result.  The parameters are $A_0=10$, $\kon=\koff=0.4$. }
\label{shnerbn}
\end{figure}

One cannot extend the above approach to compute the an exact closed-form expression for the steady-state distributions for $N>1$. However, one can make progress by recognizing that the moment equations take a particularly simple form. Consider the steady-state master equation for the case of general $N$.
We  have, in an obvious notation,
\begin{eqnarray}
0 & = & \kon P_{N-1} - 2\koff P_N  + \frac{d}{dE} \Big( \big[  (B_0+A_p)  E -A_p  \big] P_2\Big) \nonumber \\
0 & = & - j\kon P_{j} - (N-j)\koff P_{j}  +  (N-j+1)\kon P_{j-1} + (j+1)\koff P_{j+1} \nonumber\\
&\ &\qquad\qquad\qquad {} +  \frac{d}{dE} \left( \left[  \left(B_0+\frac{jA_p}{N} \right) E -\frac{jA_p}{N} \right] P_j\right) \qquad\qquad j=2\ldots N-1\nonumber \\
0 & = & - N \kon P_0 + \koff P_1+ + \frac{d}{dE} \Big(  B_0  E   P_0 \Big)
\label{dichoteq_N}
\end{eqnarray}
Because of the form of these equations, we can get a closed linear system for the moments
$z_n \equiv \int dE P_n (E) E$:
\begin{equation}
0 = - (N-j)\kon z_j - j\koff z_j  +  (N-j+1)\kon z_{j-1} + (j+1)\koff z_{j+1}  - \left(B_0+\frac{jA_p}{N}\right)  z_j - \frac{jA_p}{N}  \; \Pi _j
\label{moment_N}
\end{equation}
where $\Pi _n \equiv \int dE P_n (E)$ are just the binomial occupation probabilities and $z_{-1} = z_{N+1} \equiv 0$. This $(N+1)\times(N+1)$ linear system can immediately be solved for the $z$'s for any given $N$.  In Fig. \ref{shnerbn}, we plot the results obtained by solving these equations to determine $\langle E \rangle = \sum _{n=0}^{N} z_n$.

For  large $N$, the distribution $\Pi_j$ becomes highly peaked around its mean, $j=N\kon/(\kon+\koff)$, and so does $z_j$.  Thus, to leading order, we can approximate
$jA_p/N$ by its mean, $A_0\equiv A_p \kon/(\kon + \koff)$, allowing us to solve the resulting system,
\begin{equation}
z_j \approx  z_j^{(0)} = A_0/(A_0 + B_0) \Pi_j = E_0 \Pi_j
\end{equation}
Thus, $\langle E \rangle = E_0$, and moreover the mean value of $E$, conditioned on the value of the input $j$ is in fact independent of $j$, so that adaptation become perfect for large $N$.  To investigate the finite $N$ effects, we expand around $z_j$, $z_j =z_j^{(0)} + \Delta_j$ using the fact that for all the important modes, $j/N- \kon/(\kon + \koff)$ is small,
of order $O(N^{-1/2})$.  To solve the resultant system, 
we can approximate it by an ODE for $\Delta(y)$ thought of as a function of the continuous variable $y= \sqrt{N}(n/N-\kon/(\kon+\koff)$. This calculation is presented in Appendix 2, and leads to the result 
\begin{equation}
 \langle E\rangle \approx E_0 - \frac{A_p^2 B_0\sigma^2}{N(A_0 + B_0)^2(A_0 + B_0 + \omega)} 
 \end{equation}
 where $\omega = 1/\tau = \kon+\koff$.
 Thus,
 \begin{equation}
1 -  \langle E \rangle/E_0 \approx  \frac{A_0 B_0(1-\bar{x})}{N\bar{x}(A_0 + B_0)(A_0 + B_0 + \omega)} 
\label{mean-correction}\end{equation}
Thus again $\langle E \rangle/E_0$ starts at 1 for $B_0=0$ but here  falls initially with slope $(1-\bar{x})/(N\bar{x}(A_0+\omega))$.  Note that this is quite different from the $B=0$ slope for $N=1$, which was independent of $A_0$.  Again, $\langle E \rangle / E_0$ has a minimum as a function of $B_0$, here at
$B_0=A_0\sqrt(1 + \omega/A_0)$, with a value at the minimum of $(1-\bar{x})/(N\bar{x} (1 + \sqrt{1+\omega/A_0})^2)$.  Thus again the suppression is largest for  $\koff \ll \kon$.  Note that for small $\kon$, the suppression appears to grow without bound, although it is in fact bounded above by unity, indicating that the smaller $\bar{x}$, the larger $N$ has to be in order for the large $N$ results to be valid.  Nevertheless, the suppression is still strong in this limit.

\section{Gaussian Model \textendash\ Large $N$ Limit}

We have seen that our approximate binomial model  becomes Gaussian in the large $N$ limit.  It is interesting to check that this agrees with the
appropriate limit of the Gaussian model for small noise.  For small noise, the Gaussian model gives a finite answer, since our criterion is automatically satisfied.  Indeed,
upon expanding Eq. \ref{mean-eq} to linear order in $\sigma_S^2$ and performing the integral we get
\begin{equation}
\langle E \rangle \approx E_0\left[1 - \sigma_S^2\frac{\DI\DA^2\tau\SI(\DA-\DI)(\DA\DI+S_0\SI+S_0\SA)(\SA S_0\DI\tau+S_0\SI\DA\tau+\DI^2\DA\tau+\DI\DA^2\tau+\DA\DI)}{(\SA S_0\DI+S_0\SI\DA+\DA\DI^2)(\DI\tau+1)(\DA+\DI)(\SA S_0\DI+S_0\SI\DA+\DA^2\DI)(\SA S_0\DI\tau+S_0\SI\DA\tau+\DA\DI)(\DA\tau+1)S_0}\right]
\end{equation}
and then taking $\SA$ and $\DA$ to $\infty$ with $\SA/\DA = A_0/\bar{x}$
and taking $\SI$ and $\DI$ to zero with $\SI/\DI = B_0/\bar{x}$, and setting $S_0=\bar{x}$, $\sigma_S^2=\sigma^2/N$, we indeed reproduce our large $N$ result, Eq. \ref{mean-correction}.  In addition, the general result confirms that the suppression is  maximized in our distinguished limit $\DA \gg \DI$.

\begin{figure}
\includegraphics[width=0.7\textwidth]{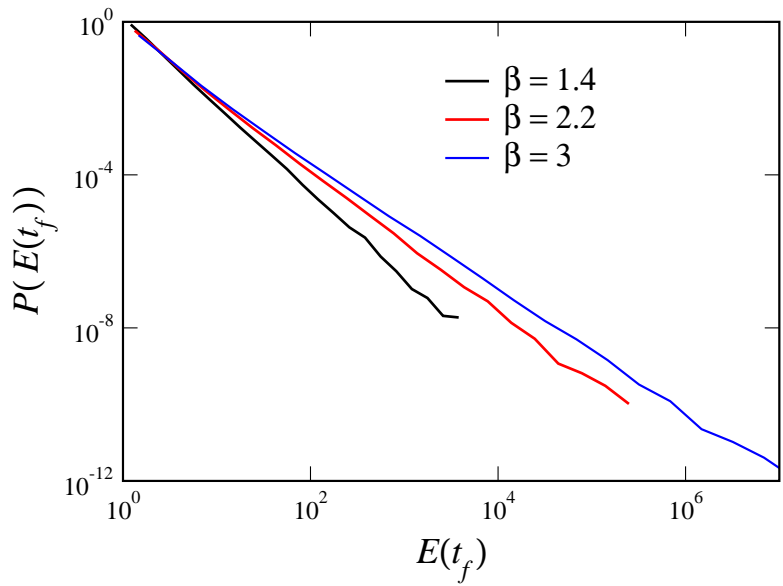}
\caption{(color online) Gaussian Model: Probability Distribution Function for $1-E(t_f)$, conditioned on $E(t_f)<0$, plotted in log-log scale, for $\SA=1.4$, $2.2$ and $3.0$.  The tails approach straight lines for large $|E(t_f)|$, indicating asymptotic power-law distributions with approximate exponents of $2.2$, $1.8$ and $1.6$, respectively.  $N=0.5, t_f=20$ All other parameters are as in Fig. 1.  For each data set, $10^6$ runs were performed, yielding 146,000, 203,000 and 234,000 data points satisfying $E(t_f)<0$, respectively and the data was binned logarithmically using 25 bins.}
\label{powers}
\end{figure}

Thus, the Gaussian model is perfectly acceptable in the small noise limit.  One can ask if there is way to extend it beyond this limit.  Clearly just increasing the noise amplitude leads to problems, as we have seen.  We have seen in Fig. \ref{comp} that the problem is not restricted just to noise levels bigger than the critical value.  Rather, for this case of large noise, the Gaussian answer is not accurate even when we are not close to the critical value. 
The problem is of course the already demonstrated large negative excursions. For small or intermediate noise levels, the problematic negative excursions of $A$ and $B$ are actually quite rare.  To understand this better, consider the distribution of $E(t)$, for some given $t$.  For short times this is well-behaved, but one exceed the time-scale of the $E$ dynamics, the distribution develops a power-law tail for large negative $E(t)$; this can be seen in Fig. \ref{powers}.  If $\SI<\SI_c$, the exponent of this distribution is greater that 2 in magnitude, and so
the first moment is finite. Of course, there is a range for which the first moment is finite but the second and higher are already divergent. Since the exact formula shows that there is no divergence at finite $t$, there must be a cutoff in the power-law tail at some extremely large value,  which however is very difficult to see from the numerics~\cite{eli}. For $\SI>\SI_c$, on the other hand, the power decreases and the first moment diverges as well  (subject to the same extremely large cutoff). As the noise level decreases, the above picture still hold.  The probability of $E(t)<0$, however, decreases exponentially with $N$.  Thus, while for $\SI>\SI_c$,
the first moment diverges,  it becomes exponentially more difficult to see this in a simulation at intermediate noise levels.  For small noise, it is well-nigh impossible. Thus, simulating the Gaussian model will, for small and intermediate noise give perfectly physical answers, which is, however, the incorrect answer for the true ensemble average, which is dominated by extremely rare huge events. Of course, if one wishes to rely on simulations, one can simulate directly the binomial noise model.

This line of reasoning leads to an alternate approach for extending the Gaussian model,
 loosely motivated by the successful use of simple cutoffs in regularizing reaction-diffusion equations which arise from the large $N$ limit of Markov processes and which overemphasize the growth at small concentrations by missing the essential role of particle number discreteness~\cite{brener-dla,perelson,evolution}. (In fact, it has been recently proven that adding a cutoff to the Fisher equation~\cite{derrida,sander,pechenik} exactly reproduces the anomalous front velocity correction  for asymptotic large $N$~\cite{proof}). Here, we cut off the integral in Eq. \ref{mean-eq} at the point where either the integrand becomes (unphysically) negative or (unphysically)  increasing as $u$ increases. Fig. \ref{comp} presents results for  $N=1$ showing that this method not only prevents blow-up (by construction) but also does a decent job in capturing the true answer. It is clear that this method becomes more and more accurate as $N$ increases, since for large $N$ the unphysical neglected integrand is exponentially small in $N$ (even though diverging as $u \to \infty$); this can be seen in Fig. \ref{comp10} .  This is reminiscent of the situation we encountered with the Gaussian simulations, where the part of the distribution that has diverging mean has exponentially small probability.

\section{Dynamic Behavior}
\begin{figure}
\includegraphics[width=0.7\textwidth]{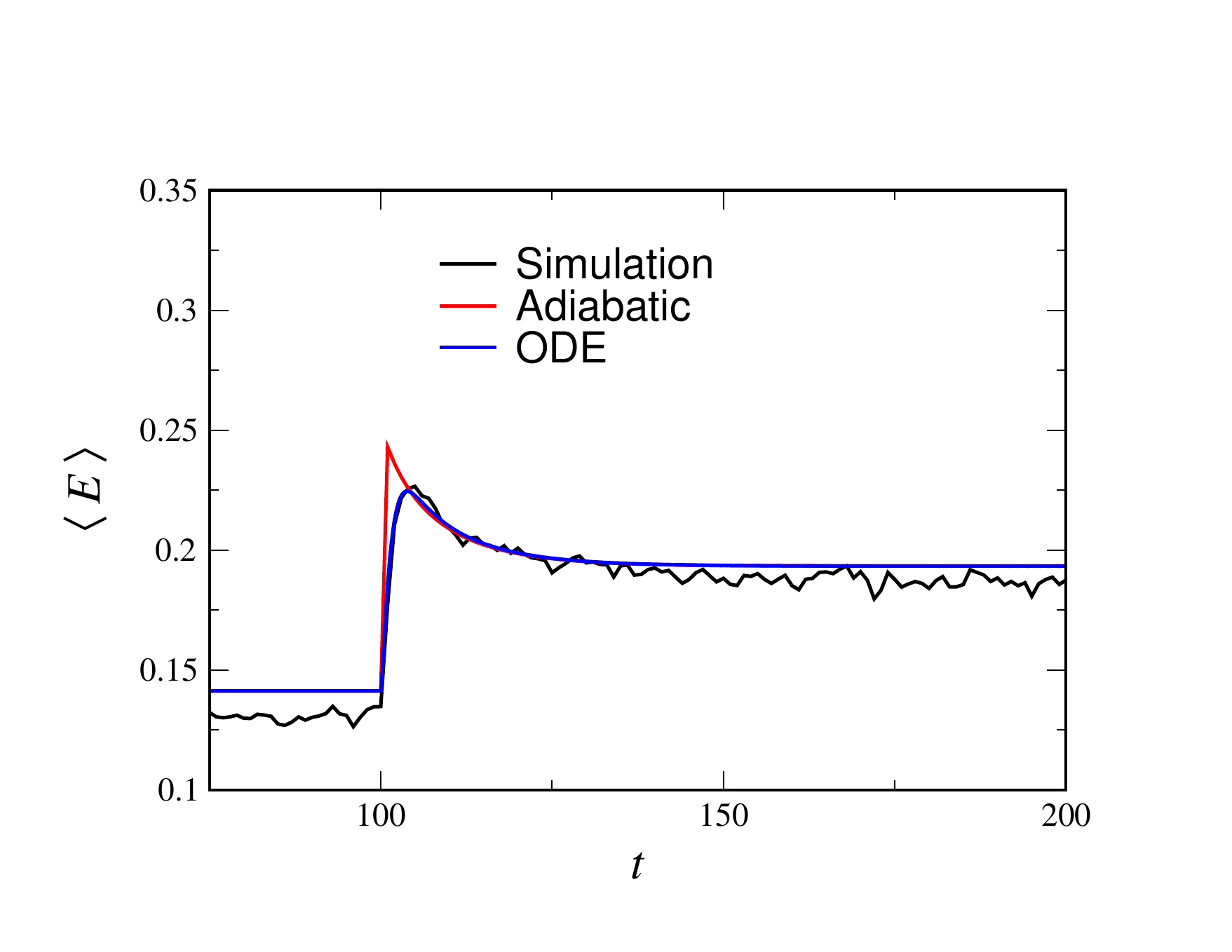}
\caption{The ensemble-averaged $\langle E(t) \rangle$ for the binomial model when $c_0$ is suddenly increased by a factor of 2 at $t_p=100$.  The average over $10^4$ runs is denoted by the curve labeled Simulation.  The adiabatic approximation where the equilibrium value of $E$ is shown with $B_0$ taken to be its ensemble average $B0(t) = B_0' - (B_0' - B_0(t_p))e^{-\delta(t-t_p)}$, and the other parameters post-pulse are taken to be their new values is shown by the trace labeled Adiabatic.  The numerical solution of the moment equations with $B_0(t)$ taken as above and $\Pi_1(t)=\bar{x}' - (\bar{x}'-\bar{x})e^{-\delta (t-t_p)}$
is marked by ODE.  The parameters are $N=1$, $\alpha=100$, $\beta=1$, $\gamma=10$, $\delta=0.1$, and $\koff=0.8$.  The initial value of $\kon=0.08$, and post-pulse $\kon=0.16$.  Thus $\bar{x}=1/11$ and $\bar{x}'=1/6$.  The deterministic equilibrium value of $E$ is 1/2, both pre- and post-pulse.}
\label{pulse}
\end{figure}

Up to this point, we have focused exclusively on the steady-state properties of the model.  It is also interesting to investigate the dynamic behavior of the model, especially since in the deterministic limit adaptation implies that the steady-state behavior is independent of the environment, and the only nontrivial
behavior is the transient response of the system to changes in the input.  In our distinguished limit where the $A$ dynamics is fast and the $B$ dynamics is slow, we can get a complete picture of the dynamics.  Imagine that the input signal suddenly changes to a new value.  The $A$ field will immediately respond
to this change.  The $B$ field will only relax slowly to the change.  As before, the relatively fast noise is averaged over, and so the $B$ dynamics can be taken to be a deterministic exponential relaxation to its new equilibrium value $B_0'$, on the slow time scale.  Thus, we can consider the changes in $B$ as adiabatic, with the system in quasi-equilibrium with the current value of $B_0$.  Thus, the value of $E$, averaged over the relatively quick fluctuations will be that given by our above solution for $\langle E \rangle$ with $B_0=B_0(t) = B_0' + (B_0'-B_0(t_p))e^{-\delta (t-t_p)}$, where $t_p$ is the time of the pulse, and $\kon$ is given by its new value, with $A_p$ and $\koff$ unchanged.  We see this plotted for the case of $N=1$ in Fig. \ref{pulse}, together with the full simulation, averaged over $10^4$ runs.  We see that this treatment captures very well the dynamics on the long $1/\delta$ time scale of the $B$ dynamics.  We see that for this strong a noise, the transient response of the system is swamped by the shift in the equilibrium value of $\langle E \rangle$. One can in fact do better by solving the time-dependent moment equations, with $\Pi_1(t) = \bar{x}' - (\bar{x}'-\bar{x})e^{-\delta (t-t_p)}$.  This resolves the initial period of the rise in $\langle E(t) \rangle$, during which the noise equilibrates to its new statistics.  This is also shown in the figure.

\section{Summary}
In summary, we have shown that the Gaussian noise model fails in principle beyond a critical value of $\SI$, the strength of the inhibitor production.  This renders a full analytic treatment impossible.  The cause of this can be traced to the interaction of the signaling nonlinearity with the fact that Gaussian noise does not respect the constraint that chemical signals must be positive. One can get rid of this effect by linearizing the reaction  equation, but this completely eliminates the possibility of investigating the extent to which the perfect adaptation seen in the deterministic limit is undermined by signal stochasticity. 

Even though it cannot be solved exactly, one can make analytic progress for the most interesting range of parameters, namely where the activation is fast and the inhibition slow, as compared to the noise time scale . These can be used as a guide to determining the actual deviation of $\langle E \rangle$ from its signal-independent mean-field value. Also, it is possible to devise a cutoff procedure which accurately predicts the results of the binomial model  for intermediate and large $N$

\acknowledgments
We thank Yu-hai Tu for stimulating discussions. This work
is supported by the NSF Center for Theoretical Biological Physics Grant No. PHY-0822283.  The work of N. Bostani is supported in part by the National Natural
Science Foundation of China, under grant numbers 11133002, 10821061, 11050110113, and by a research fellowship for international young scientists of the Chinese Academy of Sciences.

\appendix
\section{Appendix I:  Gaussian Noise}
We present here the formulas for $\Phi$ and $\Delta E(u)$ appearing in the expression for the average $E$ for the Gaussian noise model.
\begin{eqnarray}
\Phi(t) &=&  -\frac{\tau\SA}{\DA^2(1-\DA^2\tau^2)}\left(\frac{\SA}{\DA} + \frac{2\SI}{\DA + \DI}\right)\left(1 - e^{-\DA t}\right)\nonumber\\
&\ &{}\qquad - \frac{\tau\SI}{\DI^2(1-\DI^2\tau^2)}\left(\frac{\SI}{\DI} + \frac{2\SA}{\DA+\DI}\right)\left(1 - e^{-\DI t}\right) \nonumber\\
&\ &{} \qquad - \tau^2\left(\frac{\SA}{\DA + 1/\tau} + \frac{\SI}{\DI + 1/\tau}\right)\left(\frac{\SA}{\DA - 1/\tau} + \frac{\SI}{\DI - 1/\tau}\right)\left(1 - e^{-t/\tau}\right) \nonumber\\
&\ &{} \qquad + \tau\left(\frac{\SA}{\DA} + \frac{\SI}{\DI}\right)^2 t 
\end{eqnarray}
\begin{eqnarray}
\Delta_E(t) &=& \frac{\tau\SA}{\DA^2(1-\DA^2\tau^2)}\left(1-e^{-\DA t}\right) + \frac{2\tau\SI}{\DI(\DA+\DI)(1-\DI^2\tau^2)}\left(1-e^{-\DI t}\right) \nonumber\\
&\ &{} \qquad + \frac{\tau}{\DA+1/\tau}\left(\frac{\SA}{\DA-1/\tau} + \frac{\SI}{\DI-1/\tau}\right)\left(1-e^{-t/\tau}\right)
\end{eqnarray}
\section{Appendix II: Large $N$ Limit}
We begin by writing the exact equation satisfied by $\Delta_j \equiv z_j - E_0 \Pi_j$:
\begin{equation}
0 = - (N-j)\kon \Delta_j - j\koff \Delta_j  +  (N-j+1)\kon \Delta_{j-1} + (j+1)\koff \Delta_{j+1} - \left(B_0+\frac{jA_p}{N}\right)  \Delta_j + \frac{j-N\bar{x}}{N} A_p (1-E_0 ) \Pi _j
\end{equation}
where $\bar{x} \equiv \kon/(\kon+\koff)$.
To proceed, we transform this difference equation into an ODE in terms of the variable  $y\equiv (j - N\bar{x})/\sqrt{N}$, and writing $\Delta_j = N^{-1}z^{(-1)}(y) +
N^{-3/2} z^{(2)}(y)$. To leading order in $N$ we get ($\omega\equiv \kon+\koff)$:
\begin{equation}
\omega \left( \frac{d}{dy} yz^{(1)}(y) + \sigma^2 \frac{d^2}{dy^2} z^{(1)}(y) \right) - (A_0 + B_0) z^{(1)} = -\frac{A_p B_0}{(A_0 + B_0)\sqrt{2\pi  \sigma^2}}y e^{-y^2/2\sigma^2}
\end{equation}
the solution of which is
\begin{equation}
z^{(1)}(y) = \frac{A_p B_0}{(A_0+B_0+\omega)(A_0+B_0)\sqrt{2\pi\sigma^2}} ye^{-y^2/2\sigma^2}
\end{equation}
This correction reflects the breakdown of perfect adaptation for finite $N$, but due to its asymmetry, does not lead to a correction in $\langle E \rangle$.  To obtain the leading order correction for this quantity, we have to go to next order:
\begin{equation}
\omega\left( \frac{d}{dy} yz^{(2)} + \sigma^2 \frac{d^2}{dy^2} z^{(2)} \right) - (A_0 + B_0) z^{(2)} = yA_p z^{(1)} - \frac{\omega(1-2\bar{x})}{2}\frac{d^2}{dy^2}yz^{(1)}+ \frac{y^2(1-2\bar{x})}{2\sigma^2\sqrt{2\pi\sigma^2}}\left(1 - \frac{y^2}{3\sigma^2}\right)\frac{A_p B_0}{A_0 + B_0}e^{-y^2/2\sigma^2}
 \end{equation}
 The solution is then
 \begin{equation}
 z^{(2)}(y)
 =  -\frac{A_p B_0 e^{-y^2/2\sigma^2}}{(A_0 + B_0)(A_0 + B_0 + \omega)\sqrt{2\pi\sigma^2}}\Bigg[ \frac{ A_p\sigma^2}{(A_0+B_0+2\omega)} \left(\frac{y^2}{\sigma^2} + \frac{2\omega}{A_0+B_0}\right) + (\bar{x}-\nicefrac{1}{2})\left(\frac{y^4}{3\sigma^4} - \frac{y^2}{\sigma^2} \right)\Bigg]
 \end{equation}
 The second term does not contribute to $\langle E \rangle$, interestingly enough, and 
 \begin{equation}
 \langle E\rangle \approx E_0 - \frac{A_p^2 B_0\sigma^2}{N(A_0 + B_0)^2(A_0 + B_0 + \omega)} 
 \end{equation}


\begin{thebibliography}{noise}
\bibitem{ras-footnote} A GEF (guanine nucleotide exchange factor) converts the inactive form of RAS (RAS-GDP) to its active form by changing the GDP to a GTP. A GAP (GTPase Activating Protein) catalyzes the reverse reaction.
\end{thebibliography}
\end{document}